\documentclass[%
superscriptaddress,
nofootinbib,
 amsmath,amssymb,
 aps,prd,
 twocolumn
]{revtex4-2}

\usepackage{graphicx,subfigure}
\usepackage{dcolumn}
\usepackage{bm}
\usepackage{natbib} 
\usepackage[colorlinks=true,citecolor=blue,linkcolor=blue,urlcolor=blue]{hyperref}
\setcitestyle{square, numbers, comma, sort&compress} 
\usepackage{slashed}

\newcommand{\bs}{\boldsymbol}

\begin{document}


\title{
Quark anomalous magnetic moment in an extreme magnetic background from pertutbative QCD
}

\author{Eduardo S. Fraga}
\email{fraga@if.ufrj.br}
\affiliation{Instituto de Física, Universidade Federal do Rio de Janeiro, Caixa Postal 68528, 21941-972, Rio de Janeiro, RJ, Brazil}

\author{Leticia F. Palhares}
 \email{leticia.palhares@uerj.br}
 \affiliation{Universidade do Estado do Rio de Janeiro, Instituto de F\'\i sica-Departamento de F\'\i sica Te\' orica, Universidade do Estado do Rio de Janeiro, Rua S\~ ao Francisco Xavier 524, 20550-013, Rio de Janeiro, RJ, Brazil}
 
\author{Cristian Villavicencio}
 \email{cvillavicencio@ubiobio.cl}
\affiliation{Centro de Ciencias Exactas and Departamento de Ciencias B\'asicas,
Facultad de Ciencias, Universidad del B\'\i o-B\'\i o, Casilla 447, Chill\'an, Chile}%


\begin{abstract}
We compute the one-loop QCD correction to the photon-quark-antiquark vertex in an extremely strong magnetic background, i.e., one in which $\sqrt{eB}$ is much larger than all other mass scales. We resort to the lowest-Landau level approximation, and consider on shell fermions. We find that the total magnetic moment is such that the anomalous magnetic moment contributes to the electric part $\sim iE_3\sigma_3$, the other contributions to $\boldsymbol{B}\cdot\boldsymbol{\sigma}$ and $i\boldsymbol{E}\cdot\boldsymbol{\sigma}_\perp$ being completely suppressed. We show results for the anomalous magnetic moment of quarks up and down, scaled by their values in the vacuum, as a function of the external magnetic field, considering dressed gluons and a renormalization scale $\mu_\mathrm{\tiny \overline{ MS}}=\sqrt{|eB|}$.
\end{abstract}

\maketitle


\section{Introduction}
\label{introduction}

The original one-loop calculation of the anomalous magnetic moment \cite{Schwinger:1948iu}, together with the spectacular agreement with experimental data that followed refined higher-order perturbative calculations (see Ref. \cite{Aoyama:2019ryr} for a review), represents a major landmark of theoretical physics.

The one-loop correction to the anomalous magnetic moment (AMM) of a fermion that moves in the presence of an external magnetic background will, however, deviate from the original prediction by Schwinger of $\alpha/2\pi$ \cite{Schwinger:1948iu,Schwinger:1949ra}, where $\alpha \approx 1/137$ is the QED fine-structure constant. This was already known by Schwinger, since what is currently known as the {\it Schwinger phase} appears in each fermion propagator \cite{Schwinger:1951nm}. The functional dependence of this correction on the magnitude of the external magnetic field has first been addressed almost half a century ago \cite{Baier:1975uj,Baier:2000yv}.

Motivated by the ultrastrong magnetic fields that can be achieved in current experiments at the Large Hadron Collider (LHC), we compute the one-loop perturbative QCD correction to the photon-quark-antiquark vertex in an extremely strong magnetic background, i.e., one in which $\sqrt{eB}$ is much larger than all other mass scales. Here $e$ is the fundamental electric charge and $B$ is the magnetic field strength. For such high fields, we can resort to the lowest-Landau level (LLL) approximation, and consider on shell fermions.

The AMM of quarks in the presence of an external magnetic field can be related to the phenomenon of magnetic catalysis \cite{Gusynin:1995nb,Shovkovy:2012zn,Miransky:2015ava} and, in the presence of a hot plasma, might play a role in the so-called chiral magnetic effect \cite{Fukushima:2008xe}. The ultrahigh magnetic fields that can be produced in high-energy heavy ion collisions \cite{Kharzeev:2007jp,Skokov:2009qp,Deng:2012pc} bring the question of whether the AMM might prove to be experimentally relevant in this physical setting. Therefore, effects from the AMM have been under investigation in quark and hadronic matter for over two decades \cite{Bicudo:1998qb,Ferrer:2009nq,Strickland:2012vu,Fayazbakhsh:2014mca,Ferrer:2015wca,Mao:2018jdo,Chaudhuri:2019lbw,Coppola:2019uyr,Mei:2020jzn,Chaudhuri:2020lga,Ghosh:2020xwp,Xu:2020yag,Ghosh:2021dlo,Farias:2021fci,Chaudhuri:2021skc,Kawaguchi:2022dbq,Mao:2022dqn,Chaudhuri:2022oru,Tavares:2023oln}

Usually, estimates of magnetic corrections to the fermion AMM are based on the Schwinger ansatz \cite{Schwinger:1951nm}, which relates the AMM with the tensor spinor structure of the self-energy. On the other hand, not much has been done in obtaining the AMM from radiative corrections to the fermion-photon coupling. The AMM has been calculated from the fermion-photon vertex, including magnetic field corrections, using the Schwinger proper-time method for {\it low} magnetic fields \cite{Lin:2021bqv}.

In the paper at hand, conversely, we explore the magnetic effects on the {\it quark} AMM in the presence of a {\it very intense} magnetic background. We find that the total magnetic moment is such that the AMM in the presence of a very strong magnetic background contributes to the electric part $\sim iE_3\sigma_3$, which should not be confused with an anomalous electric dipole moment \cite{Schwartz:2014sze}.
The other AMM contributions to $\boldsymbol{B}\cdot\boldsymbol{\sigma}$ and $i\boldsymbol{E}\cdot\boldsymbol{\sigma}_\perp$ are completely suppressed.

We also show results for the anomalous magnetic moment of quarks up and down, scaled by their values in the vacuum, as a function of the external magnetic field, considering dressed gluons and a renormalization scale $\mu_\mathrm{\tiny \overline{ MS}}=\sqrt{|eB|}$. We consider two scenarios: one with a constant gluon mass,  $m_g=0.3$\,GeV, and one with the gluon mass extracted from the one-loop correction to its polarization tensor in the presence of a large external magnetic field.

The paper is organized as follows. In Sec.\,\ref{AMM-formalism} we present the general formalism. First, we outline the computation of the anomalous correction in QED. Then, we compute the AMM of QCD in a strong magnetic background, with special emphasis on the QCD correction to the vertex triangular diagram that
contributes to the structure function $F_2$. In Sec.\,\ref{results} we present and discuss our results for the QCD correction to the AMM as a function of the magnetic field in a few phenomenological scenarios. Finally, Sec.\,\ref{summary} contains our summary and outlook.

\section{Anomalous magnetic moment: formalism}
\label{AMM-formalism}

\subsection{Outline of AMM in one-loop QED}
\label{AMM-textbook}

In order to set the notation and define the main physical quantities, we start with a brief discussion of well-known textbook results for the anomalous magnetic moment in QED (see, e.g., Ref. \cite{Schwartz:2014sze}).

The Dirac equation 
\begin{equation}
[i\slashed{D}-m]\psi=0
\end{equation}
provides, at tree level, the fermion-antifermion-photon vertex $e_q \overline{u}(p')\gamma^\mu u(p)$, where $m$ is the spinor mass, $e_q$ its electric charge ($e_q=-e$ for the electron), and $D_{\mu}=\partial_{\mu}+i e_q A_{\mu}$ the covariant derivative. Here, we use the Weyl representation for the Dirac gamma matrices in Minkowski space.
The equation of motion naturally exhibits the AMM when written in quadratic form, so that charged spinors in an electromagnetic field $F_{\mu\nu}$ satisfy
\begin{equation}
\left[ D^2 +m^2 - \frac{e_q}{2}F_{\mu\nu}\sigma^{\mu\nu}\right]\psi =0 \,,   
\end{equation}
where $[D_\mu,D_\nu]=-ie_qF_{\mu\nu}$, and $\sigma^{\mu\nu}=\frac{i}{2}[\gamma^\mu,\gamma^\nu]$. 
%
%
%
%
%
%
%
One-loop corrections bring about other structures, with terms proportional to $(p'+p)$ and $(p'-p)$. 
In this case, one can resort to the Gordon identity
\begin{equation}
\overline{u}(p')(p'^\mu+p^\mu)u(p)= 
\overline{u}(p')\left[  2m\,\gamma^\mu     -i\sigma^{\mu\nu}(p'_\nu-p_\nu)       \right]  u(p)\,,
\end{equation}
where the fermions are on shell. 
The photon momentum $q=p'-p$ can be recognized in the previous equation, so the contribution to the fermion-photon vertex $\gamma_\mu$ is interpreted as a correction to the fermion charge, while the term $\sigma_{\mu\nu}q^\nu$ is identified with the AMM.

At any loop order, the vertex is constrained by Lorentz symmetry, the Dirac equation for spinors, $\slashed{k}_i u(k_i)=m\, u(k_i)$, the Ward identity, and the Gordon identity. So, one can write it in the following standard form:
\begin{equation}
{\cal M}^\mu=e_q \overline{u}(p') \left[ F_1\left(\frac{q^2}{m^2}\right)\gamma^\mu + F_2\left(\frac{q^2}{m^2}\right)\frac{i\sigma^{\mu\nu}}{2m}q_\nu \right]u(p) \,.
\end{equation}
On one hand, one can see that the function $F_1$ encodes charge renormalization, which brings a scale dependence to $e_q$. On the other hand, $F_2$ has the structure of a magnetic moment, and indeed
\begin{equation}
g=2 \left[ 1+ F_2\left(\frac{q^2}{m^2}\right)  \right]\,.
\end{equation}

\begin{figure}[h]
    \centering
    \includegraphics[scale=0.6]{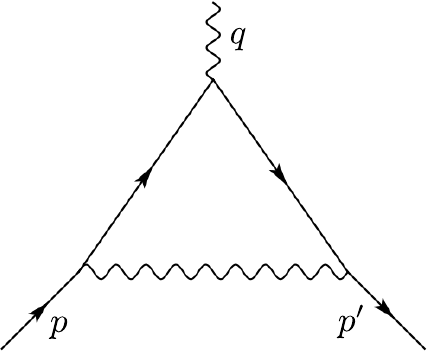}
    \caption{Vertex triangular diagram from QED that contributes to $F_2$.}
    \label{fig:triangle-photon}
\end{figure}

To one loop, only the triangular diagram of Fig.\,\ref{fig:triangle-photon} contributes to $F_2$ and, consequently, to $(g-2)$. 
Other diagrams correspond to particle-reducible corrections to the external legs and contribute only to $F_1$. 
The one-loop contribution to $F_2$ is thus given by
\begin{equation}
\begin{split}
i\delta{\cal M}^\mu=(ie_q)^3\int_k  
\,D_{\mu\nu}(k)\,
\overline{u}(p')\gamma^\nu 
S(k+p')
\\
\gamma^\mu
S(k+p)\gamma^\rho \,u(p) \,,
\label{eq.M_2}
\end{split}
\end{equation}
where the free fermion propagator in momentum space is $S(k)=i/(\slashed{k}-m+i\epsilon)$ and the gluon free propagator in momentum space in the Feynman gauge is $D_{\mu\nu}(k)=-ig_{\mu\nu}D(k)$, with $D(k)=i/(k^2+i\epsilon)$.
Here, we use the notation $\int_k \equiv \int \frac{d^4k}{(2\pi)^4}$. 

Using the standard Feynman trick of writing
\begin{equation}
\frac{1}{ABC}=\int_{xyz}\frac{2}{\left[xA+yB+zC\right]^3}\,,
\end{equation}
where we  use the following notation for the integrals in Feynman parameters,
\begin{equation}
    \int_{xyz}=\int_0^1 dx~dy~dz~\delta(x+y+z-1),
\end{equation}
and performing usual quantum field theory manipulations, yields
\begin{equation}
F_2(q^2)=\frac{\alpha}{\pi}m^2\int_{xyz} \frac{z(1-z)}{(1-z)^2m^2-xyq^2}\,.
\end{equation}
For an on shell photon ($q^2=0$), this leads to the well-known result $F_2(0)=\alpha/2\pi$, so that $g=2+\alpha/2\pi$.

\subsection{One-loop QCD correction to the photon-quark-antiquark vertex in a strong magnetic background}
\label{AMM-magQCD}

\begin{figure}[h]
    \centering
    \includegraphics[scale=0.6]{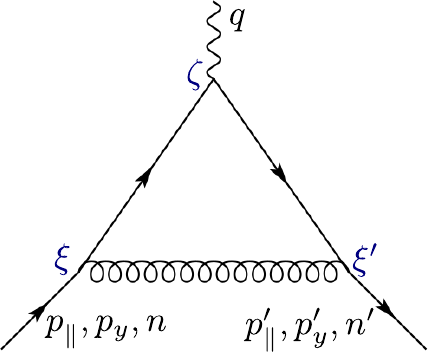}
    \caption{One-loop QCD correction to the vertex triangular diagram that contributes to $F_2$.}
    \label{fig:triangle-gluon}
\end{figure}

Let us now consider the triangle diagram depicted in Fig. \ref{fig:triangle-gluon}, encoding the first QCD correction to the quark-photon vertex.
In the vacuum, its result is analogous to the one discussed in the last section, except for the inclusion of the color structure of the $SU(N_c)$ group and the swapping of the electromagnetic coupling for the strong coupling $\alpha_s=g_s^2/4\pi$. This gives rise to a AMM correction of the following form:
\begin{equation}
    a_q^\mathrm{vac}\equiv F_2^\text{\tiny vac}(0)=\frac{\alpha_s}{2\pi}\frac{N_c^2-1}{2N_c}\,.
    \label{eq:F2vac}
\end{equation}
This kind of triangle diagrams describing corrections to some fermion-antifermion-boson interaction has been calculated in the presence of external magnetic field with different methods \cite{Ayala:2020muk,Lin:2021bqv,Villavicencio:2022gbr,Dominguez:2023bjb,Braghin:2018vyr,Braghin:2018drl,Braghin:2023ykz,Dominguez:2023tmt}.

In the presence of a strong magnetic field $\bs{B}=B \hat{\bs{z}}$, the quark fields are quantized in Landau levels. 
Here we adopt the Landau gauge representation  \cite{Parle:1987tx,Elmfors:1995gr}, with $A^{\mu}=(0,0,Bx,0)$.
So, the matrix elements for the quark field operator have the form
\begin{equation}
\langle 0|\psi (\xi)| p_\parallel,p_y,n \rangle =
 f_n (p_y,\xi_\perp) u_n(p_\parallel)\, e^{-ip\cdot\xi_\parallel}\,,
\end{equation}
where $\xi$ is a spacetime coordinate coordinate four-vector and $p^{\mu}=(p^0,p^1,p^2,p^3)$ is the four-momentum. The parallel  and perpendicular vectors are defined with respect to the magnetic field direction, $\bs{B}=B \hat{\bs{z}}$: $v_\parallel=(v^0,0,0,v^3)$ and $v_\perp=(0,v^1,v^2,0)$.
Also, $n$ denotes the index of Landau levels, $u$ is the spinor, and $f_n$ is a diagonal matrix in spinorial space related to the solution of the Dirac equation in the presence of a magnetic background \cite{Elmfors:1995gr}.
For the external photon leg, the associated 
photon field matrix element reads
\begin{equation}
\langle 0|{\cal A}_\mu (\zeta)| q \rangle = 
 \varepsilon_\mu (q)\,e^{-iq\cdot\zeta} \,,
\end{equation}
where $\zeta_{\mu}$ is a spacetime coordinate, $q_{\mu}$ is the four-momentum and
$\varepsilon_\mu$ is the polarization vector.

We use the Minkowski metric with an explicit separation of parallel and perpendicular components defined as $g_{\mu\nu}=g_{\mu\nu}^\parallel+g_{\mu\nu}^\perp$, where $g_{\mu\nu}^\parallel \equiv {\rm diag}(1,0,0,-1)$ and $g_{\mu\nu}^\perp \equiv {\rm diag}(0,-1,-1,0)$. 
This means that for any vector $v_\perp^2=-\bs{v}_\perp^2$. 
Also, we use a general notation for four-vector coordinates $(v^0,v^1,v^2,v^3)=(v_t,v_x,v_y,v_z)$.
The temporal components of the relevant external momenta are given by
\begin{align}
p_0 &=\sqrt{p_z^2+2n\ell_q^{-2}+m^2}, 
\\
p_0'    &=\sqrt{p_z'^2+2n'\ell_q^{-2}+m^2},
\\
q_0 &=|\bs{q}|,
\end{align}
where $\ell_q = |e_q B|^{-1/2}$ is the magnetic length and $m$ is the quark mass.

The one-loop vertex transfer matrix associated with the QCD correction to the quark-photon vertex is then
\begin{equation}
\begin{split}
 {\cal T} =  -e_q g_s^2 
 \sum_{l,l'}
\int_{\xi \xi' \zeta} e^{i\left( p'\cdot \xi_\parallel' - p\cdot \xi_\parallel -q\cdot \zeta \right)} \,\bar D_{\nu\rho}(\xi-\xi')
\\
\bar{u}_{n'}(p_\parallel')
f^\dag_{n'}(p_y',\xi_\perp')
\, \gamma^\nu t^a\, \bar  G_{l}(\zeta,\xi') \,
\\
\slashed{\varepsilon} \,  \bar G_{l'}(\xi,\zeta)\, \gamma^\rho t^a  \, f_{n}(p_y,\xi_\perp) u_{n}(p_\parallel) ,
\end{split}
\end{equation}
where $g_s$ is the gauge coupling constant, $t^a$ are Gell-mann matrices, and $\bar G_n$ is the fermion Green's function in coordinate space in the presence of the external magnetic field for a given Landau level $n$ \cite{Shovkovy:2012zn}.
Here, we adopt the Feynman gauge for the gluon and its propagator $\bar D_{\nu\rho}(x)$ in coordinate space can be expressed in terms of the Fourier transformation as
\begin{equation}
\bar D_{\nu\rho}(\xi-\xi')=g_{\nu\rho}
\int_{k}e^{-ik\cdot(\xi-\xi')}
D(k)\,.
\end{equation}

The quark propagator can be expressed in terms of the Fourier transformation of the parallel coordinates
\begin{equation}
\bar G_n(\xi,\zeta)=\int_{k_\parallel}
e^{-ik\cdot(\xi-\zeta)_\parallel} G_n(k_\parallel;\xi_\perp,\zeta_\perp),
\end{equation}
so that, after integrating over parallel variables, we obtain parallel momentum conservation only. Therefore, we define the resulting one-loop QCD correction to the vertex matrix as 
\begin{equation}
{\cal T} = (2\pi)^2 \delta^{(2)}(q_\parallel-p'_\parallel+p_\parallel)\,\varepsilon_\mu \,\delta{\cal M}^\mu \,,    
\end{equation}
with 
\begin{equation}
\begin{split}
\delta{\cal M}^\mu =
-e_q g_s^2 \frac{N_c^2-1}{2N_c} 
\int_{\xi_\perp\xi'_\perp\zeta_\perp k} 
e^{-ik\cdot (\xi-\xi')_\perp}  \, D(k)
\\
 \bar u'_n(p_\parallel')f_n^\dag(p_y',\xi') 
 \,\gamma^\nu\, G_{n'}(k_\parallel+p'_\parallel;\zeta_\perp,\xi'_\perp)
 \\ 
 \gamma^\mu \,G_n(k_\parallel+p_\parallel;\xi_\perp,\zeta_\perp)\,  \gamma_\nu \, f_n(p_y,\xi) u_n(p_\parallel).
\label{eq.M_mu}
\end{split}
\end{equation}
%

If we restrict our analysis to extreme magnetic fields, we can keep only the lowest-Landau level (LLL) and set $n=n'=l=l'=0$. Then, our expressions simplify to 
\begin{equation}
f_0(p_y,\xi_\perp) u_0(p_\parallel)= u_0(p_\parallel) \,\frac{e^{ip_y\xi_y -\frac{1}{2}(\xi_x/\ell_q+ sp_y\ell_q)^2}}{(\pi\ell_q^2)^{1/4}} \,,
\end{equation}
\begin{equation}
\begin{split}
G_0(k_\parallel;\zeta_\perp,\xi_\perp) = \frac{i}{2\pi\ell_q^2}S(k_\parallel)\, {\cal P}_+\qquad
\\
e^{\frac{is}{2\ell_q^2}(\xi_x+\zeta_x)(\xi_y-\zeta_y)}\, e^{\frac{1}{4\ell_q^2}(\zeta_\perp-\xi_\perp)^2}\,,   
\label{eq:G_0}
\end{split}
\end{equation}
where  $s={\rm sgn}(e_qB)$, and we have defined the projectors 
${\cal P}_\pm = \frac{1}{2} \left[ 1 \pm si \gamma_1\gamma_2\right]$ or, equivalently, ${\cal P}_\pm=\frac{1}{2}[1\pm  \Sigma_3]$, with $\Sigma_3=\rm{diag} (1,-1,1,-1)$ being the third component of the spin projection (cf. also Ref. \cite{Shovkovy:2012zn}).
The Schwinger phase corresponds to the first  exponential in the second line of Eq.\,\eqref{eq:G_0}.

The LLL matrix can be separated in parallel and perpendicular integrals in the following form:
\begin{align}
   \delta {\cal M}^{\mu}_\text{\tiny LLL} &=
    \int_{k_\perp} {\cal F}(p_y,p_y',k_\perp)\, {\cal G}^\mu(p_\parallel,p'_\parallel,k_\perp^2)\,,
    \label{eq.MLLL}
\end{align}
with
\begin{widetext}
\begin{align}
    {\cal F} &= \frac{ e_q g_s^2}{3\ell_q^5\pi^{5/2}}\int_{\xi_\perp \xi'_\perp \zeta_\perp} 
    \exp \Bigg\{
         -i k\cdot(\xi-\xi')_\perp+ip_y\xi_y-ip'_y\xi'_y
        +\frac{(\zeta-\xi)_\perp^2}{4\ell_q^2} +\frac{(\xi'-\zeta)_\perp^2}{4\ell_q^2}
        \nonumber \\ &\qquad
        -\frac{1}{2}(\xi_x/\ell_q+ sp_y\ell_q)^2 -\frac{1}{2}(\xi_x'/\ell_q+s p_y'\ell_q)^2
        -\frac{is}{2\ell_q^2}(\xi_x+\zeta_x)(\xi_y-\zeta_y) -\frac{is}{2\ell_q^2}(\zeta_x+\xi_x')(\zeta_y-\xi'_y)
        \Bigg\},
    \\ &\nonumber\\
         {\cal G}^\mu &=-\int_{k_\parallel}\,D(k)\,
         \bar u_0(p'_\parallel)\, \gamma^\nu 
    S(k_\parallel+p'_\parallel)\,{\cal P}_+
    \gamma^\mu
    S(k_\parallel+p_\parallel)\,{\cal P}_+
    \gamma_\nu\,u_0(p_\parallel).
    \label{eq.G_mu}
\end{align}
\end{widetext}

The integrals in ${\cal F}$ are Gaussian and can be evaluated analytically.
After a long but straightforward calculation, Eq.\,(\ref{eq.MLLL}) is reduced to 
\begin{equation}
   \delta {\cal M}^\mu_\text{\tiny LLL} = \frac{4}{3}e_q g_s^2 \int_{k_\perp}e^{\frac{1}{2}k_\perp^2\ell_q^2}{\cal G}^\mu(p_\parallel,p'_\parallel,k_\perp) \,.
\end{equation}
Notice that the dependence in $p_y$ and $p_y'$ disappears.
Because of the projectors ${\cal P}_\pm$ in Eq.\,(\ref{eq.G_mu}), the perpendicular components of ${\cal G}^\mu$ vanish and the form factors are restricted only to the parallel directions, {\it i.e.}, $\mu=0$ and $\mu=3$.
Moreover, there is only dependence on parallel momenta, so that all vectors and tensors will be reduced to their parallel components.

The detailed calculation of ${\cal G}^\mu$  
is long, but straightforward, and the full result is not particularly illuminating. 
In the end, there are two terms related to $F_2$ that are proportional to $\sigma_{\parallel}^{\mu\nu}q_\nu$ and $\gamma_1\gamma_2\,\sigma_{\parallel}^{\mu\nu}q_\nu$. 


\subsection{Implications to the AMM}
\label{AMM-result}

The AMM is better described by the polarization projections ${\cal P}_+ \,\sigma_{\parallel}^{\mu\nu}q_\nu$ and ${\cal P}_- \,\sigma_{\parallel}^{\mu\nu}q_\nu$.
Defining $ \delta{\cal M}^\mu_\text{\tiny LLL} =\delta {\cal M}^\mu_+ + \delta{\cal M}^\mu_-$ and 
using $\bar u_0(p'_\parallel)(\slashed{p}'_\parallel-m)=0$, $(\slashed{p}_\parallel-m)u_0(p_\parallel)=0$ and the Gordon identity, we obtain the following structure:
\begin{equation}
\begin{split}
    \delta{\cal M}^\mu_\pm = e_q\bar u_0(p'_\parallel)\hspace{140pt}\\ \Big[
    \delta F_1^\pm\gamma^\mu_\parallel 
   +F_2^\pm\sigma^{\mu\nu}_\parallel \frac{iq_\nu}{2m} 
    +F_3^\pm q_\parallel^\mu 
    \Big] 
    {\cal P}_\pm \,u_0(p_\parallel)\,,
    \label{eq.deltaMpm}
\end{split}
\end{equation}
so that the AMM splits in the two polarization projections. 

Here, $\delta F_1$ represents a correction to the tree level charge vertex.
All the form factors $F_i$ are functions of $q_\parallel^2/m^2$. 
$F_2$ can be recognized as the AMM contribution, but projected onto parallel components only. 
The third term $F_3$ should be interpreted as a correction to the term $\partial\cdot A_\parallel$, related to the gauge fixing.

In summary, if we consider the full magnetic moment, we have the following modification in the presence of a very large magnetic background:
\begin{equation}
   (1+a)\sigma_{\mu\nu}F^{\mu\nu} \to 
   \sum_{s=\pm}(\sigma_{\mu\nu}F^{\mu\nu}+a^s\sigma^\parallel_{\mu\nu}F^{\mu\nu}_\parallel){\cal P}_s \,,
   \end{equation}
where the parallel-project AMM contributes as $\sim iE_3\sigma_3$.
This term, however, should not be confused with an anomalous electric dipole moment \cite{Schwartz:2014sze}.
The other AMM contributions to $\boldsymbol{B}\cdot\boldsymbol{\sigma}$ and $i\boldsymbol{E}\cdot\boldsymbol{\sigma}_\perp$ are completely suppressed. Indeed, this becomes clear when the corrections are written in matrix form:
\begin{equation}
\sum_{s=\pm}a^s\sigma^\parallel_{\mu\nu}F^{\mu\nu}_\parallel{\cal P}_s = -2 i E_z\,
{\rm diag}(a^+,-a^-,-a^+,a^-)\,,
   \end{equation}
when compared to the standard anomalous term in vacuum:
\begin{eqnarray}
a\,\sigma_{\mu\nu}F^{\mu\nu} &=& 
-2a \, \begin{pmatrix}
\bs{B}\cdot \bs{\sigma}+i\bs{E}\cdot \bs{\sigma}_{\perp} & 0\\
0 & \bs{B}\cdot \bs{\sigma}-i\bs{E}\cdot \bs{\sigma}_{\perp}
\end{pmatrix}
-\nonumber\\
&&-2 i E_z\,
{\rm diag}(a,-a,-a,a)\,.
   \end{eqnarray}
Furthermore, since in general $a^+\neq a^-$, an asymmetry between the spin up and spin down corrections develops in the presence of a very large magnetic background.
   
Integrating over inner parallel momenta with the use of Feynman parameters, the AMM component of Eq.\,(\ref{eq.deltaMpm}) reads
\begin{equation}
    F_{2}^\pm(q_\parallel^2) 
    =\int_{xyz}\int_{k_\perp} \frac{ \frac{4}{3} g_s^2 \,e^{-\frac{1}{2}\bs{k}_\perp^2\ell_q^2}\,m^2 f^\pm(z)/4\pi}{\left[m^2(1-z)^2-q_\parallel^2 xy +(\bs{k}_\perp^2+m_g^2)z\right]^2}
\end{equation}
%
where
\begin{align}
    f^+(z) &= \frac{2}{3}(1+6z)\\
    f^-(z) &= z(4-z)-\frac{2}{3}.
\end{align}

To avoid anticipated infrared divergences, we simply add a mass term $m_g$ to the gluon propagator, leaving the discussion of possible physical choices and their consequences to the next subsection.

Since the form factors $F_i$ depend on $q_\parallel^2$, we can choose a frame where $q_\parallel^2=0$ if $p_z=p_z'$ or, in the ultrarelativistic case, for aligned $p_z\gg m$ and $p_z'\gg m$.
In this frame, it is easy to integrate over two of the three Feynman parameters. 
On the other hand, the perpendicular momentum integral can be evaluated in polar coordinates. 
Changing variables as $|\bs{k}_\perp|^2 = 2m^2\eta$, we obtain
\begin{align}
a_q^\pm  &\equiv F_2^\pm(0) \nonumber\\
   &=a_q^\mathrm{vac}\int_0^1dz\int_0^\infty d\eta\, \frac{e^{- (m\ell_q)^2\eta} \,f^\pm(z)(1-z)}{\left[(1-z)^2+(m_g/m)^2z+2\eta z\right]^2}.
    \label{eq:F2_B}
\end{align}
One should recall (cf. Eq. (\ref{eq:F2vac})) that $a_q^\mathrm{vac}$ contains the strong coupling $\alpha_s$ which, in principle, runs with the relevant energy scale.

\subsection{Effective mass for the gluons}
\label{sec:dressed}

\begin{figure}[h]
    \centering
    \includegraphics[scale=0.6]{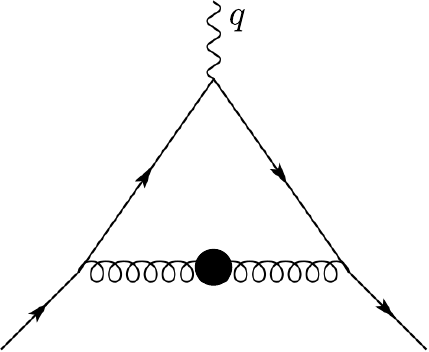}
    \caption{One-loop QCD correction to the vertex triangular diagram with a dressed gluon propagator.}
    \label{fig:triangle-dressed-gluon}
\end{figure}

We have introduced a mass term $m_g$ for the gluon in the calculation of the one-loop quark-antiquark-photon vertex in the presence of an extremely large magnetic background. This scale is necessary to regulate infrared divergences that appear in this extreme limit\footnote{It is important to note that, in this regime, the QED triangle diagram also displays the same infrared divergences and the full AMM calculation at one loop would also require an IR regularization for the photon that we do not discuss here.
}. We shall analyze the results in two different scenarios that we motivate in what follows: (i) a fixed scale and (ii) a
magnetically-dressed gluon self-energy as shown diagramatically in
Fig.\,\ref{fig:triangle-dressed-gluon}. 

First, we consider a fixed infrared scale $m_g\sim 0.3$\,GeV inspired by results for correlation functions in the nonperturbative region of QCD (and in associated pure gauge theories). For Landau and linear covariant gauges, different lattice simulations \cite{Cucchieri:2007md,Bogolubsky:2007ud,Oliveira:2010xc,Bicudo:2015rma}, Schwinger-Dyson equations  \cite{Aguilar:2008xm,Mueller:2014tea,Aguilar:2015bud} and infrared models \cite{Dudal:2008sp,Capri:2015ixa,Pelaez:2021tpq} seem to indicate the emergence of an infrared mass scale of this order.

Second, we analyze the consequences of the intense magnetic background on the gluon polarization tensor to introduce a magnetically-dressed gluon in our triangle diagram computation.
Following \cite{Fukushima:2011nu,Bandyopadhyay:2016fyd,Ayala:2018ina,Ayala:2019akk}, the gluon polarization tensor in the presence of an intense magnetic field, {\it i.e.} within the LLL approximation, is transverse in the parallel components. Hence:
\begin{equation}
    \Pi^{\mu\nu}_g(k) = \left( g_\parallel^{\mu\nu}-\frac{k_\parallel^\mu k_\parallel^\nu}{k_\parallel^2}\right)\Pi_g(k_\parallel^2,k_\perp^2) \,,
\end{equation}
where
\begin{equation}
    \Pi_g(k_\parallel^2,k_\perp^2) = \frac{\alpha_s}{\pi} N_c\sum_q\frac{e^{-\frac{1}{2}\ell_q^2\bs{k}_\perp^2 }}{\pi^2\ell_q^2}g(m_q^2/k_\parallel^2)\,.
    \label{eq:Pig}
\end{equation}
We shall adopt the chiral-limit approximation $g(m^2/k_\parallel^2)\approx g(0)=1$, so that the remaining dependence is on the perpendicular component of the momentum. Notice that the polarization factor is exponentially suppressed for $\boldsymbol{k}_\perp^2 > |eB|$. We will consider then $\Pi_g$ as a magnetically dressed gluon squared mass.

\section{QCD correction to the AMM as a function of the magnetic field}
\label{results}

Now we proceed to the numerical analysis of the behavior of the one-loop QCD correction to the AMM in the presence of a very strong magnetic background, $a_q^\pm$, compared with its vacuum counterpart, $a_q^\mathrm{vac}$, as 
defined in Eqs. \eqref{eq:F2_B} and \eqref{eq:F2vac}, respectively. In this analysis, we consider the two scenarios for the effective gluon mass discussed in the previous section: a constant gluon mass, which we take as $m_g=0.3$\,GeV, and one given by the LLL polarization tensor, $m_g^2=\Pi_g$, as defined in Eq. \eqref{eq:Pig}. In both cases we investigate what happens if we set the renormalization scale to run with the magnetic
field, $\mu_\mathrm{\tiny \overline{ MS}}=\sqrt{|eB|}$, or to a fixed value, $\mu_\mathrm{\tiny \overline{MS}}=1$\,GeV.

\begin{figure}[h]
\includegraphics[scale=0.55]{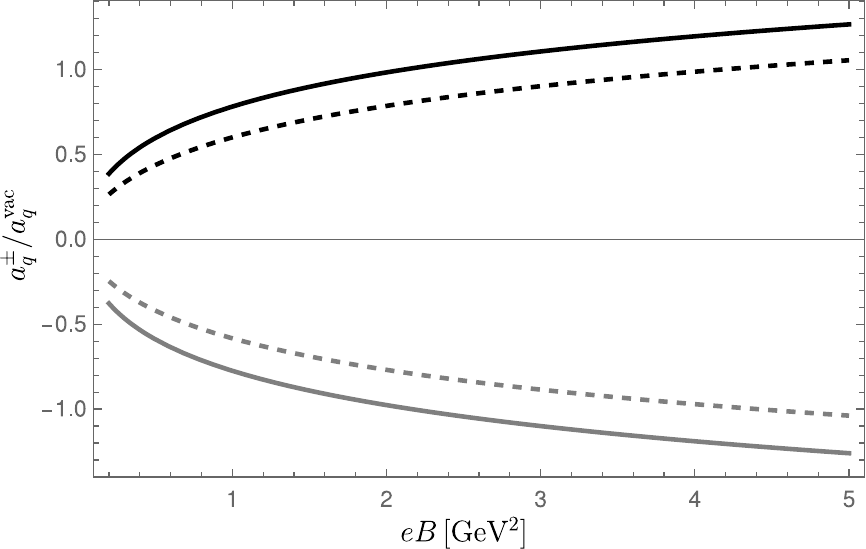}
\caption{Anomalous magnetic moments $a_u^+$ (black solid), $a_u^-$ (gray solid), $a_d^+$ (black dashed), and $a_d^-$ (gray dashed), scaled by their values in the vacuum, as functions of the external magnetic field. Here, we consider dressed gluons with $m_g=0.3$\,GeV and a renormalization scale $\mu_\mathrm{\tiny \overline{ MS}}=\sqrt{|eB|}$. 
}
\label{fig:03GeV-1GeV}
\end{figure}

\begin{figure}[h]
\includegraphics[scale=0.55]{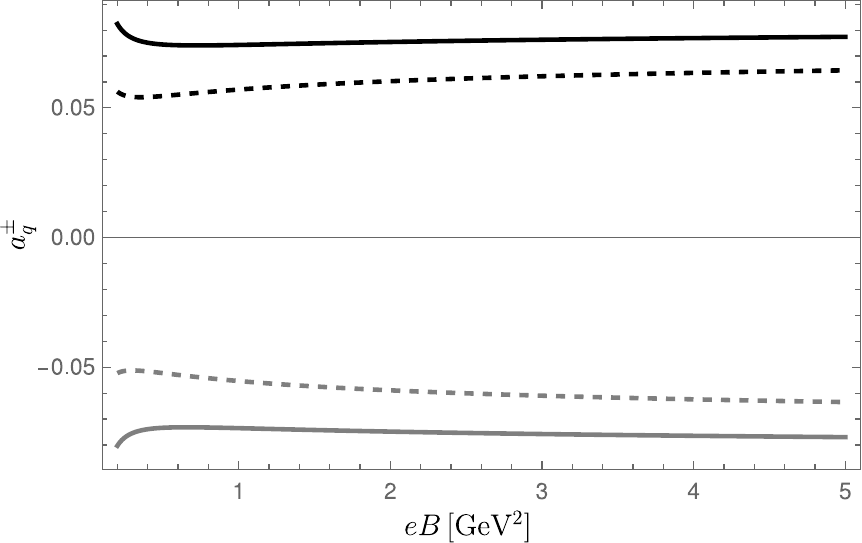}
\caption{Anomalous magnetic moments $a_u^+$ (black solid), $a_u^-$ (gray solid), $a_d^+$ (black dashed), and $a_d^-$ (gray dashed), as functions of the external magnetic field. Here we consider dressed gluons with $m_g=0.3$\,GeV and a renormalization scale $\mu_\mathrm{\tiny \overline{ MS}}=\sqrt{|eB|}$. 
}
\label{fig:03GeV-1GeV-nonscaled}
\end{figure}

Figure\,\ref{fig:03GeV-1GeV} shows the case with a fixed gluon mass, $m_g=0.3$\,GeV, and $\mu_\mathrm{\tiny \overline{ MS}}=\sqrt{|eB|}$. 
The first thing to be noticed is the change of sign of the polarization components of the AMM, being $a_q^+$ positive and $a_q^-$ negative. Moreover, we find that $a_q^-\approx -a_q^+$, within $\sim 1\%$.
One can also see that $|a_u^\pm|>|a_d^\pm|$. This happens because the quark charge, through $\ell_q$, dominates in the exponential term in Eq.\,\eqref{eq:F2_B}.
The plots present an increase in $|a_q^\pm|$ as a function of $eB$.
In fact, $|a_u^\pm|$ is greater than its corresponding vacuum value for $eB\gtrsim 2$\,GeV$^2$, and $|a_d^\pm|$ is greater than its vacuum value for $eB\gtrsim 4$\,GeV$^2$. This growth in the ratio $a_q^\pm/a_\mathrm{vac}$ is, however, somewhat misleading. It basically represents the running of $\alpha_s$ with the magnetic field in the denominator of the ratio. $a^\pm_q$ is, indeed, almost constant, as can be seen in Fig.\,\ref{fig:03GeV-1GeV-nonscaled}.

The behavior of the ratio $a_q^\pm/a_\mathrm{vac}$ for a fixed scale $\mu_\mathrm{\tiny \overline{MS}}=1$\,GeV is basically the same, since the main changes are contained in the global factor $\sim \alpha_s$. The modifications in the running quark mass have no significant impact, producing differences of $\sim 0.1\%$.

\begin{figure}[h]
\includegraphics[scale=0.55]{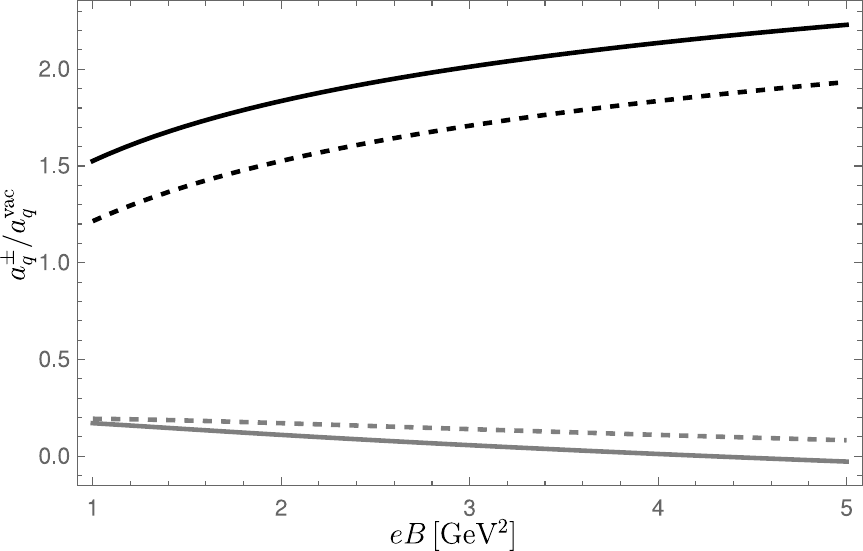}
\caption{Anomalous magnetic moments $a_u^+$ (black solid), $a_u^-$ (gray solid), $a_d^+$ (black dashed), and $a_d^-$ (gray dashed), scaled by their values in the vacuum, as functions of the external magnetic field. Here we consider dressed gluons and quarks with $m_g=0.3$\,GeV and $m_q=0.35$\,GeV. 
}
\label{fig:03GeV-035GeV-eB}
\end{figure}

\begin{figure}[h]
\includegraphics[scale=0.55]{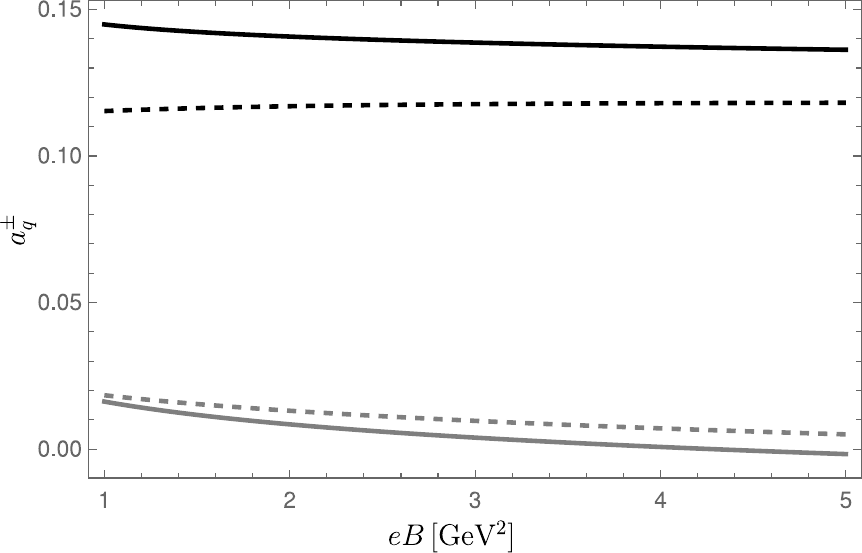}
\caption{Anomalous magnetic moments $a_u^+$ (black solid), $a_u^-$ (gray solid), $a_d^+$ (black dashed), and $a_d^-$ (gray dashed), scaled by their values in the vacuum, as functions of the external magnetic field. Here we consider dressed gluons and quarks with $m_g=0.3$\,GeV and $m_q=0.35$\,GeV, and with the  renormalization scale $\mu_\mathrm{\tiny \overline{ MS}}=\sqrt{|eB|}$.
}
\label{fig:03GeV-035GeV-eB-nonscaled}
\end{figure}

\begin{figure}[h]
\includegraphics[scale=0.55]{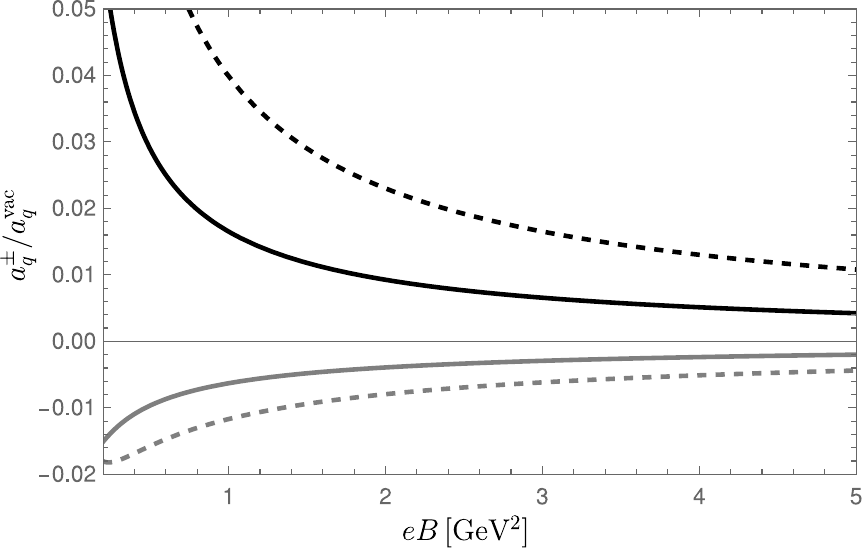}
\caption{Anomalous magnetic moments $a_u^+$ (black solid), $a_u^-$ (gray solid), $a_d^+$ (black dashed), and $a_d^-$ (gray dashed), scaled by their values in the vacuum, as functions of the external magnetic field. Here we consider dressed gluons with $m_g^2=\Pi_g$ and a renormalization scale $\mu_\mathrm{\tiny \overline{ MS}}=\sqrt{|eB|}$.
}
\label{fig:03GeV-eB}
\end{figure}

Since we consider an effective nonperturbative gluon mass of $0.3$ GeV, we also show the analogous results for a constituent quark mass of $m_q=0.35$ GeV in Figs.\ref{fig:03GeV-035GeV-eB} and \ref{fig:03GeV-035GeV-eB-nonscaled}. The larger quark mass brings no remarkable qualitative modification, yielding approximately a shift in the positive direction.

A totally different outcome occurs when we consider the dressed gluon mass as given by the LLL polarization tensor, $m_g^2=\Pi_g$, and a renormalization scale $\mu_\mathrm{\tiny \overline{ MS}}=\sqrt{|eB|}$. Contrary to what happens for a constant gluon mass, the absolute value of the AMM now diminishes as the magnetic field increases, as illustrated in Fig. \ref{fig:03GeV-eB}. This behavior is expected since $\Pi_g$ grows with the magnetic field. If we consider a fixed renormalization scale, $\mu_\mathrm{\tiny \overline{ MS}}=1$\,GeV, the plot looks very similar. However, the difference is larger than in the case of a fixed gluon mass. The difference between the two cases increases $\sim 10\%$ for $eB\sim 5$\,GeV$^2$ but stabilizes at higher values.


The strong magnetic background effects on the electron AMM are, of course, analogous. One  has only to substitute the running strong coupling by the electromagnetic coupling and set the global factor $(N_c^2-1)/2N_c\to 1$ in Eq.\,\eqref{eq:F2vac}, besides considering one flavor with charge $e$ and with $N_c\to 1$ in Eq.\,\eqref{eq:Pig}.

\begin{figure}[h]
\includegraphics[scale=0.55]{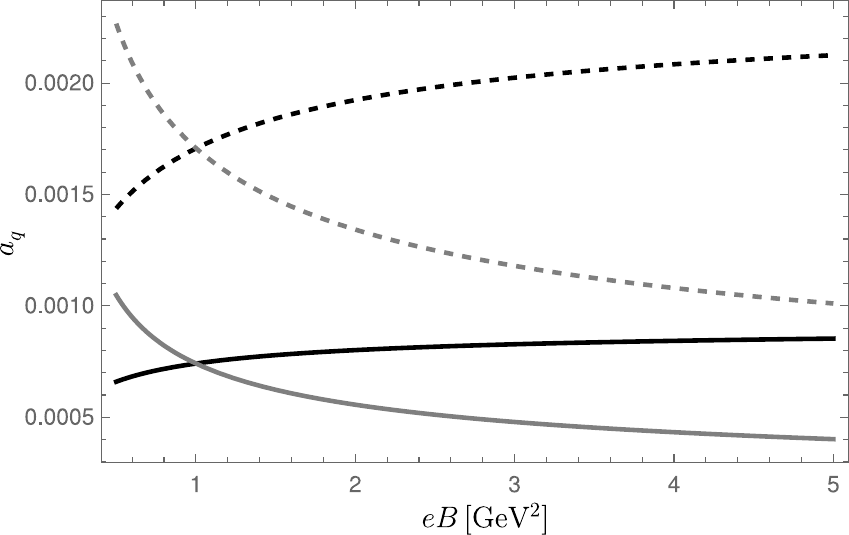}
\caption{Total anomalous magnetic moments $a_u=a_u^++a_u^-$ (solid), and $a_d=a_d^++a_d^-$ (dashed), as functions of the external magnetic field. Here we consider dressed gluons with $m_g = 0.3$\,GeV, current quark masses and the renormalization scales $\mu_\mathrm{\tiny \overline{ MS}}=\sqrt{|eB|}$ (gray) and $\mu_\mathrm{\tiny \overline{ MS}}=1$\,GeV (black).
}
\label{fig:03GeV-total_eB}
\end{figure}

\begin{figure}[h]
\includegraphics[scale=0.55]{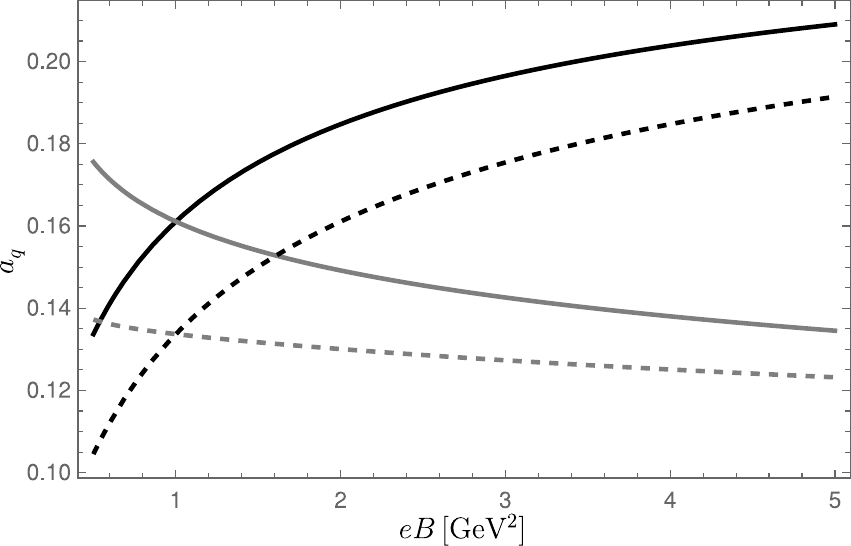}
\caption{Total anomalous magnetic moments $a_u=a_u^++a_u^-$ (solid), and $a_d=a_d^++a_d^-$ (dashed), as functions of the external magnetic field. Here we consider dressed gluons with $m_g=0.3$\,GeV, $m_q=0.35$\,GeV and the renormalization scales $\mu_\mathrm{\tiny \overline{ MS}}=\sqrt{|eB|}$ (gray) and $\mu_\mathrm{\tiny \overline{ MS}}=1$\,GeV (black).
}
\label{fig:03GeV-035GeV-total_eB}
\end{figure}

\begin{figure}[h]
\includegraphics[scale=0.55]{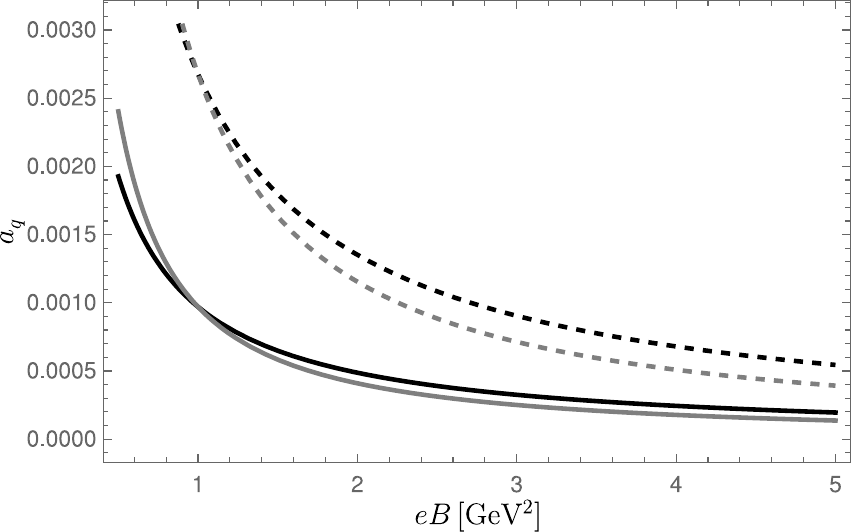}
\caption{Total anomalous magnetic moments $a_u=a_u^++a_u^-$ (solid), and $a_d=a_d^++a_d^-$ (dashed), as functions of the external magnetic field. Here we consider dressed gluons with $m_g^2 = \Pi_g$, current quark masses and the renormalization scales $\mu_\mathrm{\tiny \overline{ MS}}=\sqrt{|eB|}$ (gray) and $\mu_\mathrm{\tiny \overline{ MS}}=1$\,GeV (black).
}
\label{fig:Pi_g-total_eB}
\end{figure}


The total anomalous magnetic moments, $a_u=a_u^++a_u^-$ and $a_d=a_d^++a_d^-$, as functions of the external magnetic field are shown in Figs. \ref{fig:03GeV-total_eB}, \ref{fig:03GeV-035GeV-total_eB} and \ref{fig:Pi_g-total_eB} for the three cases considered above. As previously, remarkable qualitative modifications occur only when we consider the dressed gluon mass as given by $m_g^2=\Pi_g$ and a renormalization scale $\mu_\mathrm{\tiny \overline{ MS}}=\sqrt{|eB|}$, where the absolute value of the total AMM diminishes as the magnetic field increases.

\section{Summary and outlook}
\label{summary}

In this paper we computed the part relevant to the quark AMM of the one-loop QCD correction to the photon-quark-antiquark vertex in an extremely strong magnetic background, i.e., one in which $\sqrt{eB}$ is much larger than all other mass scales. This justified the use of the lowest-Landau level approximation. The solution exhibits infrared divergences. These can be tamed by attributing an effective mass to gluons that acts as an infrared regulator, inspired by what is expected to happen in the infrared sector of strong interactions \cite{Mena:2023mqj}. In this vein, we considered two standard scenarios: one with a constant gluon mass, $m_g=0.3$\,GeV, which sits in the ballpark of $\Lambda_{QCD}$, and one in which the gluon mass comes from the one-loop correction to its polarization tensor in the presence of an external magnetic field in the LLL approximation, $m_g^2=\Pi_g$.
In both cases we investigated what happens if we set the renormalization scale to run with the magnetic field, $\mu_\mathrm{\tiny \overline{ MS}}=\sqrt{|eB|}$, or to a fixed value, $\mu_\mathrm{\tiny \overline{ MS}}=1$\,GeV.

It has been previously shown that the AMM is induced by chiral symmetry breaking, being related to a nonperturbative contribution \cite{Ferrer:2008dy,Chang:2010hb}. In our calculation, the AMM also vanishes in the chiral limit. However, in our case we consider only perturbative vertex corrections. So, when spontaneous symmetry breaking occurs, the generation of an AMM is indeed expected. The previous references consider Schwinger-Dyson equations which in fact incorporate nonperturbative effects. This could be introduced here through operator product expansion as a nonperturbative effect, which is out of the scope of the present analysis.

Considering the total magnetic moment, we found that it is such that the anomalous magnetic moment contributes to the electric part $\sim iE_3\sigma_3$, the other contributions to $\boldsymbol{B}\cdot\boldsymbol{\sigma}$ and $i\boldsymbol{E}\cdot\boldsymbol{\sigma}_\perp$ being completely suppressed. We found that the anomalous contribution can be naturally separated into two polarization projections obtained from ${\cal P}_\pm= \frac{1}{2}(1\pm is \gamma_1\gamma_2)$, or equivalently $\frac{1}{2}(1+\Sigma_3)$.

We presented results for the behavior of the anomalous magnetic moment of quarks up and down, scaled by their values in the vacuum, as a function of the external magnetic field, for the choices of gluon mass and renormalization scale discussed above. We found that the case with a fixed gluon mass exhibits an increase of the ratio $a^\pm/a_\mathrm{vac}$, while this observable in the case with gluon mass given by $m_g^2=\Pi_g$ decreases as the magnetic field increases. The latter setup, which seems to be closer to a more physical description of the gluon mass, shows a strong suppression, {\it i.e.}, the quark AMM apparently vanishes for extremely large magnetic fields. 

We also found a asymmetry between spin up and spin down components of the anomalous correction, which might be more promising in searching possible experimental observables. 


A natural extension of the treatment presented here would be considering the case with other Landau levels in the external lines of the triangular diagram. This would bring information on the interplay between different levels in the computation of the quark anomalous magnetic moment.

One could also incorporate effects from a thermal or dense medium to the framework discussed here. To do that, one would need to consider hot and dense magnetic QCD \cite{Blaizot:2012sd,Ayala:2021nhx,Fraga:2023cef,Fraga:2023lzn,Dominguez:2023bjb,Dominguez:2023tmt} in the computation of the photon-quark-antiquark vertex. The thermal case could be relevant for high-energy heavy-ion collisions, and even play a role in the chiral magnetic effect scenario, since the quark AMM could destroy the fermion zero mode in the presence of a magnetic field. The case at high densities is of interest in magnetars, where ultrahigh magnetic fields can also be achieved. Although in this case the effect on the equation of state seems to be minor \cite{Ferrer:2015wca}, it contributes to increase the level of pressure anisotropy \cite{Strickland:2012vu}. Finally, a calculation of the quark AMM within lattice QCD in the presence of a strong magnetic background would provide a clean benchmark.


\bigskip

\begin{acknowledgments}
This work was partially supported by CAPES (Finance Code 001), CNPq, FAPERJ, INCT-FNA (Process No. 464898/2014-5) and ANID/FONDECYT under Grant No. 1190192. 
\end{acknowledgments}






%

\end{document}